\documentclass[twocolumn,showpacs,amsmath,amssymb,prb,floatfix]{revtex4}

\usepackage{graphicx}
\usepackage{dcolumn}
\usepackage{bm}

\begin{document}

\title{Relaxation and Glassy Dynamics in Disordered Type-II Superconductors}

\author{Michel Pleimling and Uwe C. T\"auber}
\affiliation{Department of Physics, Virginia Tech, Blacksburg, VA 24061-0435}

\date{\today}

\begin{abstract}
We study the non-equilibrium relaxation kinetics of interacting magnetic flux 
lines in disordered type-II superconductors at low temperatures and low 
magnetic fields by means of a three-dimensional elastic line model and Monte 
Carlo simulations.
Investigating the vortex density and height autocorrelation functions as well 
as the flux line mean-square displacement, we observe the emergence of glassy 
dynamics, caused by the competing effects of vortex pinning due to point 
defects and long-range repulsive interactions between the flux lines.
Our systematic numerical study allows us to carefully disentangle the 
associated different relaxation mechanisms, and to assess their relative impact
on the kinetics of dilute vortex matter at low temperatures. 
\end{abstract}



\maketitle

\section{Introduction}
\label{sec:level1}

In this paper, we report an investigation of the non-equilibrium relaxation
kinetics in the vortex glass phase of layered disordered type-II 
superconductors. 
Since Struik's original investigations,\cite{Struik} many glassy systems have
been found to exhibit physical aging phenomena, which have attracted 
considerable interest during the past decades.\cite{MPBook} 
Recently, it has been realized that glass-like relaxation and aging can in fact
be found in many other systems.\cite{Henkel09, Cugliandolo02, Henkel08}
Glassy materials feature extremely long relaxation times which facilitates the
investigation of aging phenomena in real as well as in numerical experiments.
Our definition of physical aging here entails two fundamental properties:
First, we require relaxation towards equilibrium to be very slow, typically
characterized by a power law decay, observable in a large accessible time 
window $t_{\rm mic} \ll t \ll t_{\rm eq}$; here $t_{\rm mic}$ denotes an 
appropriate short microscopic time scale, whereas $t_{\rm eq}$ is the much 
larger equilibration time for the macroscopic system under consideration.
Second, a non-equilibrium initial state is prepared such that the kinetics is
rendered non-stationary; thus, time-translation invariance is broken, and 
two-time response and correlation functions depend on both times $s$ and 
$t > s$ independently, not just on the elapsed time difference $t - s$.
In this context, $s$ is often referred to as waiting time, and $t$ as 
observation time.
In addition, in the limit $t \gg s$ many aging systems are characterized by the
emergence of dynamical scaling behavior.\cite{Henkel09}

The physics of interacting vortex lines in disordered type-II superconductors 
is remarkably complex and has been a major research focus in condensed matter 
physics in the past two decades. 
It has been established that the temperature vs. magnetic-field phase diagram 
displays a variety of distinct phases.\cite{Blatter}
A thorough understanding of the equilibrium and transport properties of vortex 
matter is clearly required to render these materials amenable to optimization 
with respect to dissipative losses, especially in (desirable) high-field 
applications.
Investigations of vortex phases and dynamics have in turn enriched condensed
matter theory, specifically the mathematical modeling and description of 
quantum fluids, glassy states, topological defects, continuous phase 
transitions, and dynamic critical phenomena.
An appealing feature of disordered magnetic flux line systems is their 
straightforward experimental realization which allows direct comparison of
theoretical predictions with actual measurements.
The existence of glassy phases in vortex matter is well-established 
theoretically and experimentally.\cite{Blatter, Banerjee}
The low-temperature Abrikosov lattice in pure flux line systems is already
destroyed by weak point-like disorder (such as oxygen vacancies in the 
cuprates).
The first-order vortex lattice melting transition of the pure system 
\cite{Nelson} is then replaced by a continuous transition into a 
disorder-dominated vortex glass phase.\cite{FisherM, Feigelman, Nattermann}
Here, the vortices are collectively pinned, displaying neither translational 
nor orientational long-range order.\cite{Divakar}
In addition, there is now mounting evidence for a topologically ordered 
dislocation-free Bragg glass phase at low magnetic fields or for weak 
disorder;\cite{Nattermann, Giamarchi, Kierfeld, FisherD, Banerjee}
and an intriguing intermediate multidomain glass state has been proposed. 
\cite{Menon}

Unambiguous signatures of aging in disordered vortex matter have also been 
identified experimentally: 
For example, Du {\em et al.} recently demonstrated that the voltage response of
a 2H-NbSe$_2$ sample to a current pulse depended on the pulse duration 
\cite{Du} (see also Ref.~[\onlinecite{Henderson}]).
Out-of-equilibrium features of vortex glass systems relaxing towards their
equilibrium state were studied some time ago by Nicodemi and Jensen through 
Monte Carlo simulations of a two-dimensional coarse-grained model 
system;\cite{Nicodemi} however, this model applies to very thin films only 
since it naturally disregards the prominent three-dimensional flux line 
fluctuations. 
More recently, three-dimensional Langevin dynamics simulations of vortex matter
were employed by Olson {\em et al.} \cite{Olson} and by Bustingorry, 
Cugliandolo, and Dom\'inguez \cite{BCD1, BCD2} (see also 
Ref.~[\onlinecite{BCI, IBKC}]) in order to investigate non-equilibrium 
relaxation kinetics, with quite intriguing results and indications of aging 
behavior in quantities such as the two-time density-density correlation 
function, the linear susceptibility, and the mean-square displacement.
Rom\'a and Dom\'inguez extended these studies to Monte Carlo simulations of the
three-dimensional gauge glass model at the critical temperature.\cite{Roma}

We remark that it is generally crucial for the analysis of out-of-equilibrium 
systems to carefully investigate alternative microscopic realizations of their
dynamics in order to probe their actual physical properties rather than 
artifacts inherent in any mathematical modeling.
Indeed, different mathematical and numerical representations of non-equilibrium
systems rely on various underlying {\it a priori} assumptions that can only be 
validated {\it a posteriori}.
It is therefore imperative to test a variety of different numerical methods and
compare the ensuing results in order to identify those properties that are 
generic to the physical system under investigation.
In this paper we employ Metropolis Monte Carlo simulations for a 
three-dimensional interacting elastic line model to investigate the relaxation 
behavior in the physical aging regime for systems with uncorrelated attractive
point defects.

We strive to employ parameter values that describe high-$T_c$ superconducting 
materials such as YBCO, and limit our investigations to low magnetic fields and
temperatures (typically $10$~K) in order for our disordered elastic line model 
to adequately represent a type-II superconductor with realistic material 
characteristics.
Thus we address a parameter and time regime wherein the slow dynamics is 
dominated by the gradual build-up of correlations induced by an intricate 
interplay of repulsive vortex interactions and attractive point pinning sites. 

Our work differs in crucial aspects from other recent studies 
[\onlinecite{BCD1, BCD2}]. 
As in Refs.~[\onlinecite{Nicodemi,Olson}], we consider the in type-II 
superconductors physically relevant situation where all defects serve as 
genuinely attractive and localized pinning sites for vortices, in the sense 
that they locally reduce the chemical potential (or equivalently, suppress the
superconducting transition temperature).
Our pinning potential landscape is therefore characterized by large flat regions
in space, where the vortices feel no pinning force, interspersed with small 
attractive potential minima of extension $b_0$, much smaller than the London
penetration depth $\lambda_{\rm ab}$ that sets the vortex-vortex interaction
range.
This is to be contrasted with the model used in Refs.~[\onlinecite{BCD1,BCD2}],
which is rather motivated by studies of interfaces in random environments that
are described by Gaussian distributions for the disorder 
strength.\cite{Blatter, Otterlo}
Consequently these models inevitably incorporate both attractive and repulsive
disorder, which can be viewed as mimicking a sample with a very high density of
point defects.
Alternatively, such a coarse-grained representation of pinning centers forming 
a continuous disorder landscape certainly becomes appropriate at elevated 
temperatures near $T_c$, since then the pinning range is set by the coherence
length $\xi_{\rm ab}$, which diverges as the critical point is approached.
Thus, a random medium description is best-suited for investigations of critical
phenomena, and also more easily amenable to field-theory representations.
At low temperatures, however, where $\xi_{\rm ab} \leq b_0$, our modeling of the
localized pinning centers appears more realistic, and we furthermore remark that
in this scenario repulsive defects would introduce different physics in the 
non-equilibrium relaxation and aging kinetics of vortices in superconductors, 
such as flux bunching in regions devoid of such disorder.
We therefore carefully exclude any repulsive pinning sites.
In addition, the temperatures used in Refs.~[\onlinecite{BCD1, BCD2}] appear to
be considerably higher than those studied in our present work.
Further differences can be found in the length of the vortex lines (in 
Refs.~[\onlinecite{BCD1, BCD2}] rather short lines were considered), in the 
boundary conditions, as well as in the initial preparation of the system.

We characterize the aging properties of the interacting and pinned flux lines 
through several different two-time quantities, namely the vortex 
density-density autocorrelation function, the flux line height-height
autocorrelation function, and the transverse vortex mean-square displacement. 
Investigating the influence of weak point defects, we find that the 
non-equilibrium relaxation properties of magnetic flux lines in disordered 
type-II superconductors are governed by various crossover effects that reflect
the competition between pinning and repulsive interactions. 
In the long-time limit and for not too large pinning strengths, the dynamics is
manifestly similar to that observed in structural glasses.

The structure of this paper is as follows:
In Section~\ref{sec2} we describe our model and the Monte Carlo simulation 
algorithm, and define the quantities of interest for our study. 
Our data and principal results are presented in Section~\ref{sec3}. 
In order to disentangle the different contributions to the non-equilibrium 
relaxation dynamics of our system, we first separately elucidate the effects of
attractive pinning centers and of long-range vortex-vortex interactions, before
we endeavour to analyze and understand their intriguing interplay as reflected 
in the vortex system's relaxation kinetics.
Finally, we discuss our findings in Section~\ref{sec4}, and compare them with 
other studies.

\section{Model and simulation procedure}
\label{sec2}

\subsection{Effective model Hamiltonian}

We consider a three-dimensional vortex system in the London limit, where the 
London penetration depth is much larger than the coherence length. 
We model the vortex motion by means of an elastic flux line free energy 
described in Ref.~[\onlinecite{Vinokur}], see also, e.g., 
Refs.~[\onlinecite{Sen, Rosso, Tyagi, Petaja, Bullard}]. 
The system is composed of $N$ flux lines in a sample of thickness $L$. 
The effective model Hamiltonian $H_N$ is defined in terms of the flux line 
trajectories ${\bf r}_j(z)$, with $j = 1,\ldots,N$, and consists of three 
components, namely the elastic line tension, the repulsive vortex-vortex 
interaction, and the disorder-induced pinning potential: 
\begin{eqnarray}
   H_N &=& \frac{\tilde{\epsilon}_1}{2} \sum_{j=1}^N \int_0^L \bigg\arrowvert
   \frac{d{\bf r}_j(z)}{dz} \bigg\arrowvert^2 \, dz \nonumber \\
   &+& \frac{1}{2} \sum_{i \ne j} \int_0^L V\bigl(|{\bf r}_i(z)-{\bf r}_j(z)| 
   \bigr) \, dz \nonumber \\
   &+& \sum_{j=1}^N \int_0^L V_D\bigl( {\bf r}_j(z) \bigr) \, dz .
\label{hamilt}
\end{eqnarray}
Here, the elastic line stiffness is ${\tilde{\epsilon}_1} \approx \Gamma^{-2}
{\epsilon_0} \ln (\lambda_{\rm ab}/ \xi_{\rm ab})$, with $\lambda_{\rm ab}$ and
$\xi_{\rm ab}$ denoting the London penetration depth and coherence length in 
the crystallographic ab plane (we assume the magnetic field along the c axis), 
and the anisotropy parameter (effective mass ratio) 
$\Gamma^{-2} = M_{\rm ab}/M_{\rm c}$.
The energy scale is set by $\epsilon_0 = (\phi_0 / 4\pi \lambda_{\rm ab})^2$, 
where $\phi_0 = hc / 2e$ is the magnetic flux quantum.
The expression for the elastic line energy in Eq.~(\ref{hamilt}) is valid in 
the limit $|d{\bf r}_j(z) / dz|^2 \ll \Gamma^2$. 
The purely in-plane repulsive interaction potential (consistent with the 
extreme London limit) between flux line elements is given by the modified 
Bessel function of zeroth order, $V(r) = 2 \epsilon_0 K_0(r/\lambda_{\rm ab})$,
which diverges logarithmically as $r \to 0$, and decreases exponentially for 
$r \gg \lambda_{\rm ab}$.
These vortex interactions are truncated at half the system size, which is in 
turn chosen sufficiently big such that numerical artifacts due to this cut-off 
length are minimized. 
We model point pinning centers through square potential wells with radius $b_0$
and strength $U_0$ at $N_D$ defect positions.
(For additional details, see Refs.~[\onlinecite{Bullard, Thananart}].)

\subsection{Numerical parameter values}

Our simulation parameter values~\cite{Thananart} were chosen corresponding to 
typical material parameters for YBCO as listed in appendix D of 
Ref.~[\onlinecite{Vinokur}]. 
In the following, lengths and energies are reported relative to the effective 
defect radius $b_0$ and interaction energy scale $\epsilon_0$ (using cgs 
units), and time in Monte Carlo steps (MCS), where one MCS correspond to $N L$
proposed updates of the flux line elements, with $N$ the number of flux lines 
and $L$ the number of layers.
We set the pinning center radius $b_0 = 35$ \AA, anisotropy parameter
$\Gamma^{-1} = 1/5$, and, as is appropriate at low temperatures, 
$\lambda_{\rm ab} = 34 b_0 \approx 1200$ \AA, and
$\xi_{\rm ab} = 0.3 b_0 \approx 10.5$ \AA. 
Then $\epsilon_0 \approx 1.9 \times 10^{-6}$ (in cgs units of energy / length),
and the energy scale in the line tension term becomes 
${\tilde{\epsilon}_1} \approx 0.18 \epsilon_0$. 
We systematically vary the pinning strength $U_0$ between $0$ and 
$0.2 \epsilon_0$. 
Usually, our simulations are performed at temperature $T = 10$~K, which 
corresponds to $k_B T / \epsilon_0 b_0 \approx 0.002$.
Thermal excitation energies are thus small compared to the elastic and pinning
energies, and at equilibrium we therefore expect the system to be deep in the 
glassy regime.
We do not allow for flux line cutting and reconnection processes in our 
simulations of a low-temperature and dilute vortex system.

\subsection{System preparation and simulation protocol}

We apply the standard Metropolis Monte Carlo simulation algorithm in three
dimensions with a discretized version of the above effective Hamiltonian 
(\ref{hamilt}).\cite{Bullard}
The system contains $N = 16$ flux lines in $L$ layers, with a distance $b_0$ 
between consecutive layers; and an equal number $N_D / L = 1116$ of point 
pinning centers that are however randomly distributed within the layer, with 
mean separation $\sim 9 b_0$; in comparison, the triangular vortex lattice
spacing would be $78.5 b_0$ in our dilute system.
We apply periodic boundary conditions in all three space directions, as we
are mainly interested in bulk properties. 
This is to be contrasted with Refs.~[\onlinecite{BCD1, BCD2}], where free 
boundary conditions were used along the c axis.
We have systematically changed $L$ between $10$ and $2560$ in order to 
carefully monitor finite-size effects.
The in-plane system size is $[L_x,L_y] = [ \frac{2}{\sqrt{3}} \times 
8 \lambda_{\rm ab}, 8 \lambda_{\rm ab}]$; the dimensions of the $xy$ plane were
chosen such that in the absence of disorder the system accommodates a regular 
triangular flux lattice.
In the absence of defects, we have tested that initially randomly placed 
vortices properly equilibrate to form a triangular Abrikosov flux lattice. 
We have also checked that there are no appreciable effects due to the sharp 
cut-off of the vortex interactions at $4 \lambda_{\rm ab} = L_y / 2$. 

\begin{figure}
\includegraphics[angle=0,width=\columnwidth]{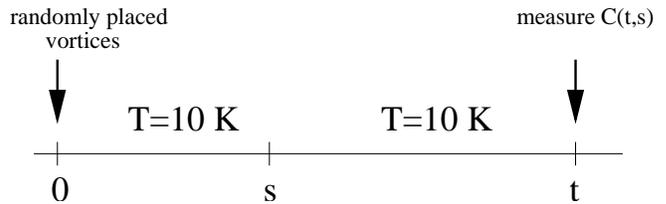}
\caption{Sketch of the measurement protocol.
   At $t = 0$, straight vortex lines are initialized far out of equilibrium by 
   placing them randomly in the system at $T = 10$~K in the presence of weak 
   point defects.
   The vortex lines are then allowed to relax for different waiting times $s$, 
   before various two-time quantities, such as the height autocorrelation 
   function $C(t,s)$, are measured.}
\label{Algorithm}
\end{figure}

In order to investigate aging phenomena in the system with uncorrelated point 
disorder, the vortices are prepared in an out-of-equilibrium state: 
Straight flux lines are initially (at $t = 0$) placed at random locations in 
the system.
The vortex lines are subsequently allowed to relax at the temperature $T=10$~K
for a duration $s$, the `waiting' time, before we start measuring two-time 
quantities for $t > s$, see Fig.~\ref{Algorithm}. 
(This is again different to Refs.~[\onlinecite{BCD1, BCD2}], where the vortex 
lines were equilibrated at high temperatures inside the vortex liquid phase 
before the subsequent quench to lower temperatures.)
Our waiting times extend up to $s = 51200$ MCS, whereas the total length of a 
simulation run is typically $t = 10 s$.

\subsection{Measured quantities}

Aging phenomena can generally be adequately characterized through the study of 
two-time quantities.
In our work we put special emphasis on a range of observables that allow us to 
rather comprehensively monitor the distinct relaxation processes in vortex
matter that originate from pinning to attractive point defects and repulsive 
interaction forces, respectively, and their intricate competitive interplay.

The height-height autocorrelation function and mean-square displacement 
represent two quantities that are routinely studied in the context of interface
fluctuations and non-equilibrium growth processes.
\cite{BCD1, BCD2, BCI, IBKC, Rothlein, Bustingorry07, Chou10} 
Separating the time-dependent position of the flux line $j$ in the $z$th layer 
into its $x$ and $y$ components, ${\bf r}_j(z,t) = \bigl( x_j(z,t), y_j(z,t) 
\bigr)$, the two-time height-height autocorrelation function can be written as
\begin{eqnarray}
  &&C(t,s) = \frac{1}{L N} \sum\limits_{z=1}^L \sum\limits_{j=1}^N 
  \big\langle x_j(z,t) x_j(z,s) \big\rangle 
  - \big\langle x(t) \big\rangle  \big\langle x(s) \big\rangle \nonumber \\
  &&\quad + \frac{1}{L N} \sum\limits_{z=1}^L \sum\limits_{j=1}^N 
  \big\langle y_j(z,t) y_j(z,s) \big\rangle - \big\langle y(t) \big\rangle  
  \big\langle y(s) \big\rangle \ ,
\end{eqnarray}
where $x(t) = \frac{1}{L N} \sum\limits_{z=1}^L \sum\limits_{j=1}^N x_j(z,t)$, 
and similarly for $y(t)$. 
The brackets $\langle \cdots \rangle$ here denote both an average over the 
noise history, i.e., over the sequential realizations of random number 
sequences, as well as a configurational average over defect distributions and 
initial positions of the straight vortex lines at the outset of the simulation
runs.  
The two-time mean-square displacement in the $xy$ planes, transverse to the
external magnetic field, can similarly be cast in the form
\begin{eqnarray}
  B(t,s) &=& \frac{1}{L N} \sum\limits_{z=1}^L \sum\limits_{j=1}^N \Bigl[ 
  \Big\langle \bigl( x_j(z,t) - x_j(z,s) \bigr)^2 \Big\rangle \nonumber \\ 
  &&\qquad\qquad + \Big\langle \bigl( y_j(z,t) - y_j(z,s) \bigr)^2 \Big\rangle 
  \Bigr] \ .
\end{eqnarray}
We remark that other related quantities that essentially contain the same 
information, are the two-time roughness function or the two-time structure 
factor.\cite{BCI, Bustingorry07}

\begin{figure}
\includegraphics[angle=0,width=\columnwidth]{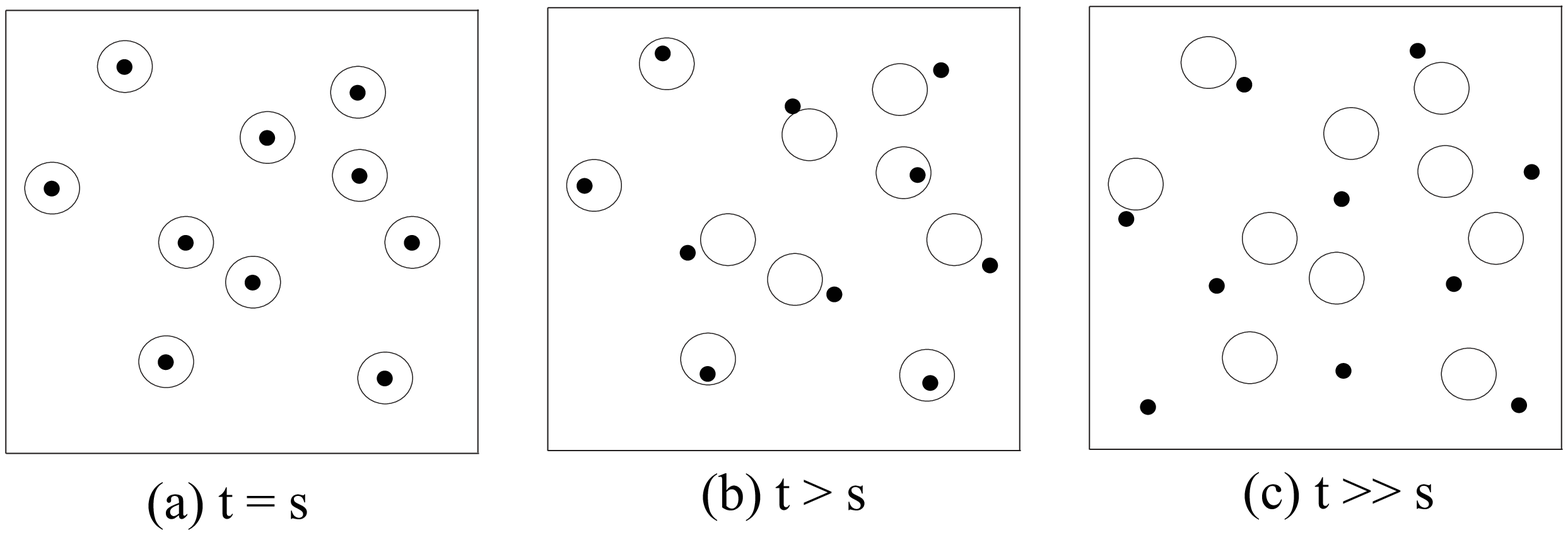}
\caption{Sketch (two-dimensional cross section) of vortex line elements (small
   solid dots) and the associated circles with radius $\alpha b_0$ (large open
   circles) at different observation times.\cite{Thananart} 
   (a) At $t = s$, each vortex line element by construction resides inside its 
   own circle and gives a count $\sum_i n_i(t=s) = N_S$ ($= 10$ here). 
   (b) At later times $t > s$, the repulsive vortex interactions cause the flux
   line elements to move away from their initial positions. 
   This results in a smaller occupation number $\sum_i n_i(t > s) < N_S$ 
   ($= 5$ in this example). 
   (c) At long times $t \gg s$, it is possible that all vortex line elements 
   have left the circle, $n_i(t \gg s) = 0$, which results in a complete 
   decorrelation. 
   For instance, the quantity $\sum_i n_i(s) n_i(t)$ in the pictures at 
   these three times is evaluated to be $10$ at $t = s$, 
   $5$ at $t > s$, and $0$ for $t \gg s$.}
\label{correlation}
\end{figure}

Unfortunately, both the height autocorrelation and the mean-square displacement
are probably not easily accessible in experiments on type-II superconductors,
except perhaps through low-angle neutron scattering.
Much better suited for an experimental study is likely the (connected) two-time
vortex density-density autocorrelation function that can formally be written as
\begin{equation}
  C_v(t,s) = \big\langle \rho({\bf r},t) \rho({\bf r},s) \big\rangle - \rho^2
  \ ,
\label{etdcor}
\end{equation}
where $\rho({\bf r},t)$ represents the local flux density per unit area at
position ${\bf r}$, with constant uniform average 
$\big \langle \rho({\bf r},t) \big \rangle = \rho$.
Following an initially random placement, the repulsive vortex interactions 
cause positional rearrangements, such that one would expect a temporal decay of
the density autocorrelation function. 
In our simulations, we realize the vortex density autocorrelation function in 
the following way:\cite{Thananart}
As before we start with randomly placed straight vortex lines at $t=0$ and let 
the system subsequently relax up to waiting time $s$. 
A density count for the vortex line elements is then generated by setting a
circular area, with a radius equal to $\alpha \, b_0$ at the location of each 
vortex line element $i \equiv (j,z)$ at $t = s$.
Typical values for $\alpha$ range from $0.05$ to $0.20$.
As time $t$ elapses, we count the number of vortex line elements still in their
circles, generating a time sequence of occupation numbers $n_i(t)$, with 
$n_i = 0$ or $1$, and $n_i(t = s) = 1$ by construction.
Due to the repulsive vortex interactions, flux line elements tend to move away
from their initial positions, whence $n_i(t > s)$ can be $0$ at a later
time if the vortex line element leaves the prescribed circle.
In the presence of pinning centers, vortex line elements will become trapped
inside the defects over a long time, causing $n_i(t > s)$ to preferentially 
remain $1$.
This quantity is then averaged over the $N_S = L N$ different vortex segments
and many distinct defect distributions and initial configurations, yielding
\begin{equation}
  C_v(t,s) = \Big\langle \frac{1}{N_S} \sum\limits_{i=1}^{N_S} n_i(s) n_i(t)
  \Big\rangle \ .
\end{equation}
Fig.~\ref{correlation} illustrates the algorithm for calculating the density
autocorrelation function.\cite{Thananart}
We have checked the results for different values of $\alpha$ and found that 
within a reasonable range the precise choice of $\alpha$ does not affect the 
results in the long-time aging regime where $t \gg s$.

\section{Relaxation processes}
\label{sec3}

In order to fully understand the non-equilibrium relaxation processes and aging
phenomena in disordered type-II superconductors at low temperatures, we found 
it imperative to carefully disentagle the dynamical contributions originating 
from the repulsive interactions between the vortex lines and from their pinning
to attractive point defects. 
We start our discussion with free flux lines, mainly in order to validate our 
code by comparing our data with the theoretically expected behavior and earlier
work. 
We will then separately consider the effects of attractive point pins and of 
the long-range repulsive vortex interactions, before we at last venture to
study the interplay of these two competing mechanisms to induce or relax
correlations in the system.

\subsection{Free elastic line}

The relaxation kinetics of a single free elastic vortex line constitutes a 
valuable benchmark to check our program as this case can be easily understood 
by recalling that in the presence of thermal noise a fluctuating interface that
tries to minimize its line tension should be described in the continuum limit 
by the linear Edwards--Wilkinson equation.\cite{Edw82} 
As the fluctuations in the transverse $x$ and $y$ directions are independent 
random variables for our free line, we expect the results for the free vortex 
to be described by the one-dimensional version of that well-known stochastic 
equation ($h$ below stands for either $x$ or $y$):
\begin{equation} 
  \frac{\partial h(z,t)}{\partial t} = \nu\, \partial_z^2 h(z,t) + \eta(z,t)\ ,
\label{eq:EW}
\end{equation}
where $\eta (z,t)$ represents a Gaussian white noise with zero mean and 
covariance $\langle \eta(z,t) \eta (z',t') \rangle = \frac{2 T}{\nu} 
\delta(t - t') \delta(z - z')$, $\nu$ is the line stiffness (equal to 
${\tilde{\epsilon}_1}$ here), and $T$ the temperature of the heat bath. 
The Edwards--Wilkinson equation, as well as a range of microscopic models 
belonging to the same dynamic universality class, has been studied extensively.
Starting from a straight line, one first observes a short-time regime with
uncorrelated fluctuations, which is rapidly replaced by a correlated 
intermediate-time interval characterized by a non-trivial power law increase of
the line roughness. 
After a crossover time that algebraically depends on the system size, this 
correlated regime finally reaches the steady-state or saturation regime.

Two-time quantities have also been studied in the context of the 
Edwards--Wilkinson equation,\cite{Rothlein, BCI, Chou10}, facilitated by the 
fact that a full analytical analysis is possible for the linear stochastic 
equation (\ref{eq:EW}). 
For example, in the correlated regime of the one-dimensional Edwards--Wilkinson
equation the following exact expression for the height-height autocorrelation 
function has been derived:\cite{Rothlein}
\begin{equation} 
  C(t,s) = C_0 s^{1/2} \left[ \left( \frac{t}{s} + 1 \right)^{1/2} - 
  \left( \frac{t}{s} - 1 \right)^{1/2} \right] \ , 
\label{eq:Corr_EW}
\end{equation}
where $C_0$ is a known constant. 
The detailed crossover properties of two-time quantities in the region between 
the correlated and saturated regimes have been carefully investigated in 
Ref.~[\onlinecite{BCI}].

\begin{figure}
\includegraphics[angle=0,width=\columnwidth]{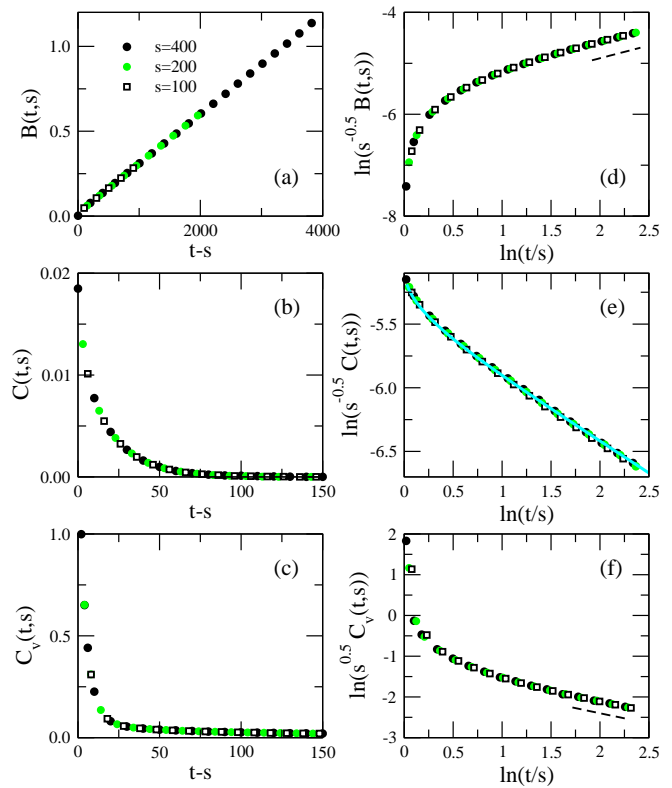}
\caption{(Color online) Various two-time quantities for the free elastic vortex
  line obtained in systems with (a-c) $L = 10$ and (d-f) $L = 2560$ planes, and
  averaged over typically 200 independent Monte Carlo simulation runs;
  (a) and (d): mean-square displacement, (b) and (e): height-height 
  autocorrelation function, (c) and (f): vortex density-density autocorrelation
  function (measured with $\alpha = 0.05$). 
  Data obtained for different waiting times are shown. 
  For $L = 10$ layers the system rapidly reaches the steady state, and 
  time-translation invariance is recovered, i.e., the two-time quantities only 
  depend on the time difference $t-s$. 
  For $L = 2560$ aging and dynamic scaling prevail throughout the simulation 
  time window.
  The full line in (e) indicates the exact expression (\ref{eq:Corr_EW}) 
  derived from the Edwards--Wilkinson equation.
  The dashed lines in (d) and (f) indicate the predicted asymptotic power laws 
  with exponents $1/2$ and $-1/2$, respectively. 
  Here and in the following figures error bars are much smaller than the 
  symbol sizes.}
\label{fig3}
\end{figure}
In Fig.~\ref{fig3}, we display our Monte Carlo simulation results for our
elastic vortex line model when both the vortex interaction and the defect 
pinning are switched off, i.e., only the first contribution in (\ref{hamilt}) 
is retained.
One immediately notices a striking difference between the behavior of a 
``thin film'' composed of only a few layers (such as $L = 10$, see 
Figs.~\ref{fig3}a--\ref{fig3}c) and ``bulk'' systems consisting of many layers 
($L = 2560$, see Figs.~\ref{fig3}d--\ref{fig3}f). 
In the former case, the system rapidly evolves into the steady-state regime, 
yielding two-time quantities that only depend on the elapsed time difference 
$t-s$. 
As a result, the transverse displacements in the $x$ and $y$ directions perform
simple random walks, as is revealed by the linear increase of the mean-square 
displacement with time, see Fig.~\ref{fig3}a. 
For the larger bulk system, the correlated regime persists throughout the 
duration of our simulations, and both waiting and observation times reside well
within that extended intermediate regime. 
This gives rise to aging and dynamical scaling: 
Time-translation invariance is broken, and all the two-time quantities display 
full-aging scaling.\cite{Henkel09} 
For each two-time observable we find the following scaling behavior (here given
for the height autocorrelation function $C(t,s)$):
\begin{equation} 
  C(t,s) = s^{-b} f_C(t/s) \ ,
\label{eq:Corr_scal}
\end{equation}
where $b$ represents an aging scaling exponent and $f_C(y)$ denotes an
associated scaling function that follows a power law decay for large arguments.
For the height-height autocorrelation we have $b = - 1/2$.\cite{Rothlein} 
In Fig.~\ref{fig3}e we explicitly compare our numerically determined scaling 
function with the expression (\ref{eq:Corr_EW}) resulting from the direct 
solution of the Edwards--Wilkinson equation (full line), and obtain perfect 
agreement. 

Summarizing, we see that the free vortex line fluctuations are indeed aptly
described by the one-dimensional Edwards--Wilkinson equation (\ref{eq:EW}). 
We also observe a strong dependence on the system's extension $L$ in the 
magnetic field direction, i.e., the vortex length: 
On the time scale of our simulations, the stationary regime is almost 
immediately reached when the system consists of only a few layers; in contrast,
for larger bulk systems aging and dynamical scaling are observed easily.
This points to quite distinct relaxation behavior in thin superconducting films
and thicker bulk samples. 
We decided to avoid the additional complications stemming from the crossover 
between the correlated and steady-state regimes in our present study, and to
rather focus on system sizes sufficiently large that no finite-size effects (no
crossover to the steady state) are observed on the accessed time scales. 
Properties of smaller systems and possible experimental consequences for thin 
superconducting films will be discussed in a separate publication.

\subsection{Pinning without interactions}

Intuitively, one anticipates pinning centers to strongly influence the thermal 
fluctuations of our elastic flux lines.
Indeed, attractive forces emanating from the pinning centers will tend to 
localize vortex segments, and thus ultimately suppress thermal fluctuations. 
Depending on the pinning strength, flux line elements will end up spending an 
appreciable amount of time close to a pinning center. 
Therefore, compared to freely fluctuating lines, a marked {\it increase} of 
correlations as function of time must be expected.

Before we proceed to analyze the influence of pinning centers in more detail, 
we need to stress that we exclusively consider attractive point defects, in 
accordance with the physics of disordered type-II superconductors in the 
low-temperature regime.
A recent study [\onlinecite{IBKC}] addressed the relaxation and aging 
properties of elastic lines subjected to a random potential, corresponding to 
both attractive and repulsive pinning centers. 
Whereas a Gaussian disorder strength distribution certainly is a good model for
disordered ferromagnets, its relevance for relaxation processes in disordered 
type-II superconductors at low temperatures is less obvious. 

\begin{figure}
\includegraphics[angle=0,width=0.7\columnwidth]{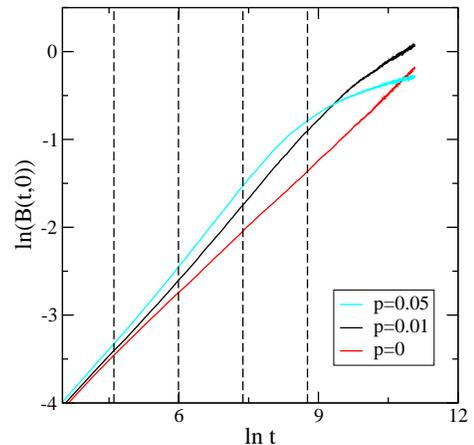}
\caption{(Color online) Mean-square displacement $B(t,0)$ vs. time $t$ (in 
  MCS) for different values of the pinning strength $p$, for systems of size 
  $L=640$. 
  In the initial time regime, the pinning centers attract the vortex segments
  as revealed by an increase of the slope of $B(t,0)$ as compared to a pure 
  system. 
  At later times the localization of the flux line elements induced by the
  pinning yields a strong decrease of this slope. 
  The data shown result from averaging over typically 100 independent runs.
  The dashed lines indicate the times $t = 100$, $400$, $1600$, and $6400$ MCS,
  see Fig.~\ref{fig5} below.}
\label{fig4}
\end{figure}
Let us start by looking at the mean-square displacement $B(t,0)$, with $s = 0$,
which gives a measure of the (squared) distance traveled by the flux line 
elements since the initial preparation of the system. 
In Fig.~\ref{fig4} we compare the behavior of a free elastic line with that of 
flux lines subject to pinning centers of various strengths 
$p = U_0 / \epsilon_0$.
The presence of attractive pins clearly gives rise to different regimes. 
The flux lines are rapidly attracted by the point defects, which yields an 
increase of the slope in the log-log plots of $B(t,0)$ vs. time $t$. 
This continues until some pinning strength-dependent crossover time at which 
the slope decreases even below the value of the free line, signifying the
confinement of localized vortex segments to the vicinity of the pins.
As one would expect, this crossover time decreases for increasing pinning 
strengths. 
For $p \geq 0.05$, $B(t,0)$ remains essentially unchanged, which indicates that
for non-interacting lines there exists a critical pinning strength above which 
thermal fluctuations are no longer sufficiently strong to allow the vortex line
elements to escape from the defects.

\begin{figure}
\includegraphics[angle=0,width=\columnwidth]{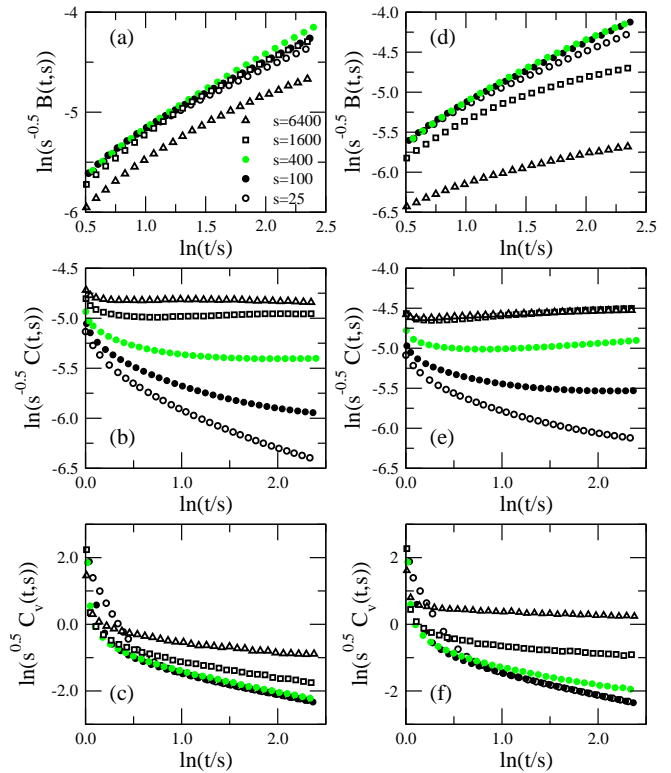}
\caption{(Color online) Various two-time quantities for non-interacting vortex 
  lines subject to attractive point defects of strengths (a-c) $p = 0.01$ and
  (d-f) $p = 0.05$: 
  (a) and (d) mean-square displacement; 
  (b) and (e) height-height autocorrelation function;
  (c) and (f) vortex density-density autocorrelation (with $\alpha = 0.05$). 
  The shown data result from averaging over typically 100 independent runs in
  systems of size $L = 640$. 
  For direct comparison with Fig.~\ref{fig3}, the data are plotted according to
  the scaling properties of free elastic lines. 
  Whereas an approximate scaling prevails for small waiting times (especially 
  for the mean-square displacement and the vortex density-density 
  autocorrelation), strong deviations emerge for larger waiting times. 
  In (e) the apparent collapse of the height autocorrelation function for the 
  largest waiting times is merely caused by the scale used in that figure.} 
\label{fig5}
\end{figure}

These different regimes also manifest themselves when two-time quantities are
considered, as seen in Fig.~\ref{fig5}, where we have plotted the data 
according to the free-line scaling behavior. 
Of course, it is not to be expected that these scaling laws remain valid when 
attractive defects are added to the system, but this representation of our data
facilitates the following discussion.
We first remark, see Figs.~\ref{fig5}a and \ref{fig5}d, that the change of the 
slope of $B(t,0)$ translates into deviations of the mean-square displacement 
$B(t,s)$ from the free-line scaling that can be readily understood.
For example, for $p=0.01$ the time intervals $[25, 250]$, $[100, 1000]$, and 
$[400, 4000]$, used to compute $B(t,s)$ for the waiting times $s=25$, $100$, 
and $400$, respectively, correspond to the time regime with increasing local 
slopes of $B(t,0)$, compare Fig.~\ref{fig4}.
This yields a shift of $\ln(s^{-0.5} B(t,s))$ to higher values.
As the crossover time of $B(t,0)$ lies inside the interval $[1600, 16000]$, the
converse behavior is observed for $s=1600$ and even larger waiting times, with 
a shift of $\ln(s^{-0.5} B(t,s))$ to lower values. 
This effect is more pronounced for larger pinning strengths, since then the 
crossover of $B(t,0)$ takes place earlier.

The strongest influence of point defects and largest deviations from free 
elastic lines are observed in the height autocorrelation function, 
Figs.~\ref{fig5}b and \ref{fig5}e. 
As the flux line elements are trapped by the pinning centers, their transverse
in-plane displacements become diminished, which leads to an increase of the 
correlations as a function of waiting time. 
In addition, the decay of $C(t,s)$ as a function of $t$ is much slower for 
larger values of $s$. 
For larger pinning strengths and long waiting times we even observe 
non-monotonic temporal evolution, as the trapped flux lines experience an 
increase of the height correlations.

Finally, the vortex density-density autocorrelation turns out to be the least 
sensitive among our two-time observables to the presence of pinning centers, 
at least for comparatively small waiting times, see Figs.~\ref{fig5}c and 
\ref{fig5}f. 
Indeed, for moderate values of $s$ one still observes the free-line scaling 
behavior; this is of course a consequence of our prescription for the 
computation of this correlation function, namely setting a circular area with 
fixed radius around every vortex line element at time $s$:
As long as only a few line elements are captured by a point defect, the scaling
of $C_v(t,s)$ remains approximately unchanged. 
Only when the majority of the vortex segments become trapped, does this 
localization induce a strong enhancement of the vortex density-density 
correlations.

\begin{figure}
\includegraphics[angle=0,width=\columnwidth]{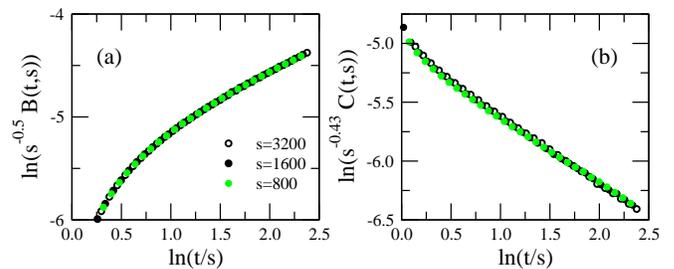}
\caption{(Color online) (a) Mean-square displacement and (b) height-height 
  autocorrelation function for non-interacting vortex lines subject to both 
  attractive and repulsive point defects with strengths drawn from the square 
  distribution $\left[ -0.01, 0.01 \right]$. 
  The shown data result from averaging over typically 100 independent runs in
  systems of size $L = 1280$. 
  The data display simple aging scaling, with different exponents for the
  two quantities.}
\label{fig6}
\end{figure}
To conclude this section, we note that the non-equilibrium relaxation physics 
is drastically different when both attractive and repulsive pinning centers are
implemented. 
As studied in Ref.~[\onlinecite{IBKC}] (see also 
Refs.~[\onlinecite{Noh09, Mon09}]) an elastic line in a random potential is 
characterized by a time-dependent correlation length that crosses over from an 
early time power law growth to an asymptotic logarithmic growth. 
Consequently, two-time quantities display an apparent simple aging scaling with
effective exponents that depend on temperature and on the randomness.
We have verified that we obtain similar results as in 
Ref.~[\onlinecite{IBKC}] when using both attractive and repulsive defects in 
our model and Monte Carlo algorithm. 
Indeed, as shown in Fig.~\ref{fig6}, the time-dependent correlation length 
gives rise to simple aging scaling of our two-time quantities, with effective 
exponents that display a dependence on temperature and on the distribution of 
the pinning strengths.
These crossover features also capture most of the relevant properties of 
disordered ferromagnets undergoing phase ordering.\cite{Par10, Cor10, Cor11} 
In Refs.~[\onlinecite{BCD1, BCD2}], disordered type-II superconductors in the 
low-temperature phase were modeled by a corresponding model with random pins 
that are either attractive or repulsive. 
However, the physical realization relevant to materials is that of purely 
attractive pins, similar to those studied in our present work. 
Yet since the properties of elastic lines strongly depend on the nature of the 
pinning centers, any conclusions regarding the non-equilibrium relaxation 
properties of disordered type-II superconductors at low temperatures that are 
inferred from models with both attractive and repulsive defects should be 
viewed with some scepticism.

\subsection{Interacting vortex lines without pinning}

\begin{figure}
\includegraphics[angle=0,width=0.7\columnwidth]{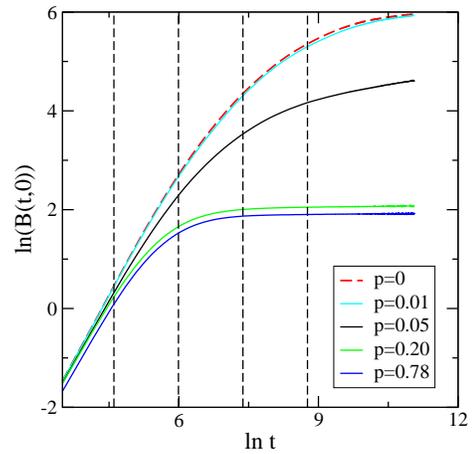}
\caption{(Color online) Mean-square displacement $B(t,0)$ vs. time $t$ (in MCS)
  for interacting vortex lines with different values of the pinning strength 
  $p$ (system size $L =640$; the data result from averaging over typically 100 
  independent runs).
  Due to the long-range interactions, the flux lines aim to maximize their
  separations, thus yielding values of $B(t,0)$ that are two orders of 
  magnitude larger than in the absence of repulsive forces, compare 
  Fig.~\ref{fig4}.
  These displacements are impeded and vortex motion eventually stopped by 
  the caging constraints of neighboring lines and pinning to attractive point 
  defects.
  The (red) dashed line displays our data for a pure system, in the absence of 
  pinning centers.
  The vertical dashed lines indicate the times $t = 100$, $400$, $1600$, and 
  $6400$ MCS.}
\label{fig7}
\end{figure}
In the absence of disorder our system composed of interacting flux lines 
evolves toward a regular triangular Abrikosov lattice. 
As we start our simulations by deposing initially straight lines at random 
positions, large displacements of the flux line elements are expected, as the 
system tries to minimize the long-range in-plane repulsive vortex interaction 
energy.
The ensuing dynamic regimes are again nicely captured by the mean-square 
displacement $B(t,0)$ which takes on values that are two orders of magnitude 
larger than in the absence of interactions, see the (red) dashed line in 
Fig.~\ref{fig7}. 
While the flux line segments experience these large displacements, $B(t,0)$ 
displays an approximate power law increase with time, with an effective 
exponent of $\approx 1.68$. 
Once the majority of vortices have reached the vicinity of their final 
equilibrium positions, the slope of $B(t,0)$ starts to gradually decrease.

\begin{figure}
\includegraphics[angle=0,width=\columnwidth]{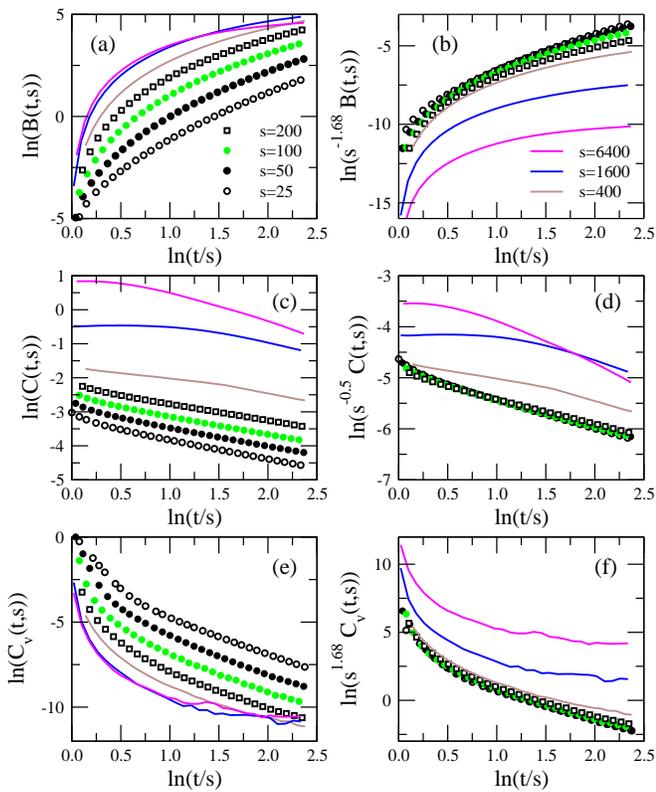}
\caption{(Color online) Two-time quantities for interacting vortices in the 
  absence of pinning centers:
  (a) and (b) mean-square displacement; 
  (c) and (d) height-height autocorrelation function; 
  (e) and (f) vortex density-density autocorrelation (with $\alpha = 0.05$). 
  Data obtained for different waiting times $s$ are shown, typically obtained
  from averaging over 500 independent simulation runs; the system size is 
  $L = 640$. 
  (a), (c), and (e) display the unscaled data in a log-log plot, whereas 
  (b), (d), and (f) show the approximate scaling observed for not too large 
  waiting times. 
  This scaling regime corresponds to the intermediate time window where 
  $B(t,0)$ exhibits a power law increase as a function of time, with an 
  (effective) exponent $\approx 1.68$, see Fig.~\ref{fig7}.}
\label{fig8}
\end{figure}
The two-time quantities reveal both the initial-time regime as well as the 
crossover at later times, see Fig.~\ref{fig8}. 
The mean-square displacement $B(t,s)$ yields a reasonably good data collapse 
for the smaller waiting times with the scaling exponent $-1.68$ that follows 
from the slope of $B(t,0)$ in that regime. 
When the observation time $t$ exceeds the crossover time, this scaling breaks 
down.
Instead the growth rate of $B(t,s)$ decreases strongly with increasing $s$, 
even resulting in a crossing of the curves for different waiting times. 
The behavior of $B(t,s)$ is mirrored by that of the 
vortex density autocorrelation function: 
For small waiting times $s$, scaling is achieved with exponent $1.68$, whereas 
for larger waiting times the decay of the correlation slows down as $s$ 
increases. 
{}From these results we infer that the vortex density-density autocorrelation 
contains essentially the same physical information as the mean-square 
displacement. 
Interestingly, the height-height autocorrelation function displays a different 
scaling for smaller waiting times, given by the scaling exponent $b = -0.5$ of 
the free line, see Figs.~\ref{fig8}c and \ref{fig8}d. 
This means that during the initial rearrangement of the vortex lines the height
fluctuations are essentially the same as for the free line. 
Only when the vortices come close to their equilibrium positions does the 
character of the correlations change, reflecting the presence of long-range
repulsive forces.

\subsection{Interacting vortex lines with pinning}

We are now ready to study the combined effects of repulsive flux line 
interactions and point defect pinning during the non-equilibrium relaxation of 
vortex matter in disordered type-II superconductors. 
We again begin by first considering the mean-square displacement $B(t,0)$, see 
Fig.~\ref{fig7}. 
Adding very weak attractive defects, e.g., with $p=0.01$, has only a very minor
effect on the time evolution of $B(t,0)$. 
Strengthening the point pins leads to a smaller rate of increase for the 
mean-square displacement, see the curve for $p =0.05$. 
At early times the flux lines are still displaced from their initial positions,
as the vortex interactions try to establish an Abrikosov lattice. 
However, at intermediate times these displacements are impeded by the defects 
that noticeably slow down vortex motion. 
As a result the system tends to a new (quasi-)equilibrium state that balances 
these two competing mechanisms. 
For even stronger pinning, the moving flux lines become rapidly trapped by the 
disorder and the system gradually freezes into a blocked configuration. 
In Fig.~\ref{fig7} this is clearly the case for both $p = 0.20$ and $p = 0.78$.

\begin{figure}
\includegraphics[angle=0,width=0.64\columnwidth]{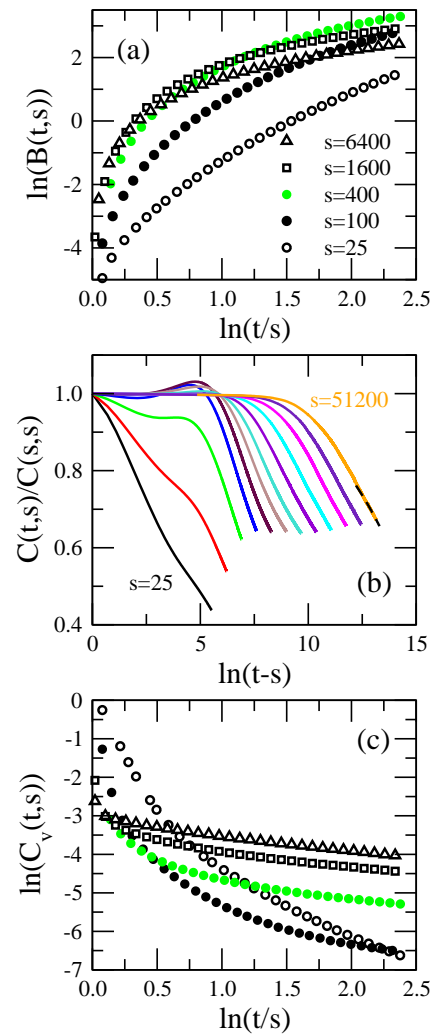}
\caption{(Color online) Two-time quantities for interacting vortices in the 
  presence of attractive pinning centers:
  (a) mean-square displacement; 
  (b) height-height autocorrelation function; and 
  (c) vortex density-density autocorrelation (with $\alpha = 0.05$). 
  The pinning strength here is $p = 0.05$, the system size $L = 640$, and the 
  data result from averaging over typically 800 independent simulation runs.
  The waiting times $s$ for the vortex density autocorrelation are the same as 
  those for the mean-square displacement. 
  For the height autocorrelation function, the waiting times range from
  $s = 25$ to $s = 51200$, the values of $s$ being doubled between consecutive 
  curves. 
  The (normalized) height-height autocorrelation function, after a crossover, 
  displays the same qualitative behavior as that encountered in structural 
  glasses. 
  For large values of $t - s$, the height autocorrelation assumes the stretched
  exponential form (\ref{eq:str_exp}) with $\beta = 0.4$, as shown by the 
  dashed line overlying the $s = 51200$ data.}
\label{fig9}
\end{figure}
The most interesting scenario naturally emerges for intermediate defect 
strengths. 
Indeed, when the pins are very weak, the rearrangement of the flux lines is 
barely affected, and all studied two-time quantities quantitatively display the
same behavior as in a pure system.
On the other hand, when the defects are too strong, the flux line elements 
remain firmly attached to the pinning centers, and a frozen configuration 
ensues. 
The non-trivial behavior at intermediate pinning strengths is studied in more 
detail in Fig.~\ref{fig9} for the case $p = 0.05$.

The data for $B(t,s)$ and $C_v(t,s)$ shown in Figs.~\ref{fig9}a and \ref{fig9}c
are readily understood by comparing them to the corresponding results for the 
pure case, see Figs.~\ref{fig8}a and \ref{fig8}e. 
The difference between these data sets is the absence of the early-time regime 
where $B(t,0)$ has an approximately constant slope, see Fig.~\ref{fig7}. 
Consequently the data with $p = 0.05$ do not allow any data collapse, not even 
for the smallest waiting times considered. 
However, this is the only noticeable difference, and the behavior for larger 
waiting times is qualitatively the same as for $p = 0$, except that the 
decrease in slope of $B(t,s)$ for larger values of $s$ is stronger when 
$p \neq 0$.

However, a completely different picture emerges for the evolution of the 
normalized height-height autocorrelation function. 
As shown in Fig.~\ref{fig9}b, for waiting times $s$ larger than a certain 
crossover value, $C(t,s)$ exhibits the typical two-step relaxation of a 
structural glass: 
An initial time-translation invariant regime, which corresponds to the 
so-called $\beta$ relaxation in glasses and only depends on the elapsed time 
difference $t-s$, is followed by a slow decay that is usually referred to as
$\alpha$ relaxation in the glass literature.\cite{Gotze}
In the long-time limit we can fit this slow decay to a stretched exponential
\begin{equation}
  f(\tau) = \exp \left[ - \left( \frac{\tau}{t_d(s)} \right)^\beta \right] \ ,
\label{eq:str_exp}
\end{equation}
with $\tau = t - s$, and a waiting-time dependent decorrelation time $t_d(s)$.
For our different waiting times we obtain a consistent value 
$\beta \approx 0.40$ for the stretching exponent in Eq.~(\ref{eq:str_exp}). 
This emergence of a characteristic two-step glass-like relaxation is very 
intriguing. 
Obviously, the flux lines do not settle into a stable microstate even after 
their lateral displacements have become strongly reduced owing to the capture
by the attractive pinning centers and the caging due to their repelling 
neighboring vortices.
Instead, as a consequence of the two competing relaxation mechanisms, 
collective dynamics and slow decorrelation sets in that yields the typical 
two-step relaxation dynamics of a glass.

We also note the intriguing shape of the normalized height autocorrelation 
function in the crossover regime.
Indeed, at intermediate waiting times $C(t,s)$ displays a strongly 
non-monotonic behavior, with a maximal value that even {\it exceeds} the value
$C(s,s)$ at $t = s$. 
This remarkable feature points to a fundamental change in the nature of the 
emerging correlations, which is due to the trapping of vortex segments in the
vicinity of the defects and the subsequent balancing of the competition between
the attractive pinning and the repulsive interactions.

\section{Discussion and conclusion}
\label{sec4}

Our three-dimensional Monte Carlo investigation of relaxation processes in 
disordered type-II superconductors has allowed us to gain a thorough 
understanding of the non-equilibrium properties of these technologically 
important materials.
We find the relaxation processes to be dominated by the interplay of two 
competing interactions, namely the pinning of the flux line elements to 
attractive point defects and the long-range mutual repulsion of the vortices.
This competition generates various crossover scenarios that we have discussed 
systematically.
The most interesting regime emerges for pinning centers of intermediate 
strength, for which we observe a distinguished two-step relaxation and a final 
slow, stretched-exponential decay of the height-height autocorrelation 
function.
This behavior is reminiscent of that encountered in structural glasses, 
clearly demonstrating that disordered type-II superconductors subject to point
defects indeed display pronounced glassy behavior at low temperatures, again
justifying the term ``vortex glass'' for this frustrated pinned low-temperature
phase.

We remark that our results are at variance with recent investigations based on 
three-dimensional London--Langevin dynamics simulations, where standard aging 
and dynamical scaling behavior of two-time quantities was 
observed,\cite{BCD1, BCD2} akin to the relaxation features of elastic lines in
a random medium.\cite{IBKC}
However, these studies, in addition to difference in sample preparation, system
size, and boundary conditions in the $z$-direction, employed a coarse-grained 
continuous random medium model of disordered type-II superconductors that 
becomes adequate near the normal- to superconducting transition, but does not
realistically capture superconducting materials at low temperatures for which
isolated defects such as oxygen vacancies always induce a local suppression of
the transition temperature and therefore constitute attractive localized 
pinning centers for vortices.
As our study shows, dynamical scaling no longer prevails for purely attractive 
point pinning centers, but instead much richer glassy relaxation dynamics sets 
in.

As already mentioned in the Introduction, it is essential for investigations of
non-equilibrium systems to study different dynamics and their algorithmic 
implementations in order to ensure that any ensuing results indeed describe 
actual physical properties of the system rather than numerical artifacts. 
We have therefore recently begun to implement corresponding London--Langevin 
dynamics simulations for our elastic line model (\ref{hamilt}) with exclusively
attractive pinning centers.\cite{Dob11} 
Our first tentative findings are in complete agreement with the Monte Carlo 
simulation results reported in this paper:
They too show the emergence of glass-like behavior, with a slow, 
stretched-exponential decay at long times. 
An in-depth analysis of this dynamics is currently in progress; this 
comparative study also aims at matching the different microscopic time scales 
implicit in Monte Carlo and Langevin dynamical simulations.

Our current study can readily be expanded in various directions. 
Our results are valid in the regime where all time-dependent length scales 
remain small compared to the size of the system. 
However, many transport and relaxation experiments are carried out on thin 
superconducting films rather than bulk samples. 
In our model, a finite (small) number of layers introduces a dominant new 
length scale that substantially changes the relaxation processes, leading to 
additional crossover features. 
We also note that other types of defects can be experimentally realized, 
ranging from parallel and splayed columnar pins to planar defects, and
combinations thereof with point disorder. 
It is an open and intriguing problem to understand how these different defect 
configurations influence the out-of-equilibrium relaxation processes in type-II
superconductors. 
A detailed understanding of the relaxation phenomena in superconducting 
materials may facilitate characterization and optimization of samples with
respect to pinning and flux transport.
Finally, in all transport applications the flux lines are driven across the 
samples by external currents, which at long times yields a non-equilibrium 
steady state replacing the thermal equilibrium state that emerges without 
drive.
Following similar lines as in the present study, one should be able to also 
analyze the relaxation properties of driven disordered type-II superconductors 
in a comprehensive manner. 
We plan to address these and related problems in the future.

\bigskip
\begin{acknowledgments}
We are indebted to Thananart Klongcheongsan for his original contributions 
during his Ph.D. dissertation work, and for preparing Figure~\ref{correlation}.
We thank Sebastian Bustingorry for useful correspondence, and Ulrich Dobramysl 
for interesting and helpful discussions.
This work was supported by the U.S. Department of Energy, Office of Basic 
Energy Sciences (DOE--BES) under grant no. DE-FG02-09ER46613.
\end{acknowledgments}


\begin{thebibliography}{99}

\bibitem{Struik}
   L.C.E.~Struik, 
   {\em Physical Aging in Amorphous Polymers and Other Materials} 
   (Elsevier, Amsterdam, 1978).

\bibitem{MPBook}
For recent overviews, see: 
   M.~Henkel, M.~Pleimling, and R.~Sanctuary (eds.), 
   {\em Ageing and the glass transition}, Lecture Notes in Physics {\bf 716} 
   (Springer, Berlin, 2007).

\bibitem{Henkel09}
   M. Henkel and M. Pleimling,
   {\em Non-Equilibrium Phase Transitions, Volume 2: Ageing and Dynamical 
   Scaling Far From Equilibrium} (Springer, 2010).

\bibitem{Cugliandolo02}
   L.F.~Cugliandolo, in:
   {\em Slow Relaxation and Non Equilibrium Dynamics in Condensed Matter}, 
   eds.~J.-L.~Barrat, J.~Dalibard, J.~Kurchan, and M.V.~Feigel'man
   (Springer, 2003).

\bibitem{Henkel08}
   M.~Henkel and M.~Pleimling, in 
   {\em Rugged Free Energy Landscapes: Common Computational Approaches in Spin 
   Glasses, Structural Glasses and Biological Macromolecules}, 
   ed.~W.~Janke, Lecture Notes in Physics {\bf 736}, 107 (Springer, 2008).

\bibitem{Blatter}
For a now classical general review, see:
   G.~Blatter, M.V.~Feigel'man, V.B.~Geshkenbein, A.I.~Larkin, and 
   V.M.~Vinokur, 
   Rev. Mod. Phys. {\bf 66}, 1125 (1994).

\bibitem{Banerjee}
   S.S.~Banerjee {\em et al.}, 
   Physica C {\bf 355}, 39 (2001).  

\bibitem{Nelson}
   D.R.~Nelson, 
   Phys. Rev. Lett. {\bf 60}, 1973 (1988);
   D.R.~Nelson and H.S.~Seung, 
   Phys. Rev. B {\bf 39}, 9153 (1989);
   D.R.~Nelson, 
   J. Stat. Phys. {\bf 57}, 511 (1989).

\bibitem{FisherM}
   M.P.A.~Fisher, 
   Phys. Rev. Lett. {\bf 62}, 1415 (1989);
   D.S.~Fisher, M.P.A.~Fisher, and D.A.~Huse, 
   Phys. Rev. B {\bf 43}, 130 (1991).

\bibitem{Feigelman}
   M.V.~Feigel'man, V.B.~Geshkenbein, A.I.~Larkin, and V.M.~Vinokur,
   Phys. Rev. Lett. {\bf 63}, 2303 (1989).
	
\bibitem{Nattermann}	
   T.~Nattermann, 
   Phys. Rev. Lett. {\bf 64}, 2454 (1990).	

\bibitem{Divakar}
For clear structural experimental evidence, see:
   U.~Divakar {\em et al.},
   Phys. Rev. Lett. {\bf 92}, 237004 (2004). 
	
\bibitem{Giamarchi}
   T.~Giamarchi and P.~Le~Doussal, 
   Phys. Rev. Lett. {\bf 72}, 1530 (1994);
   Phys. Rev. B {\bf 52}, 1242 (1995);
   Phys. Rev. Lett. {\bf 76}, 3408 (1996);
   Phys. Rev. B {\bf 55}, 6577 (1997);
   T.~Klein, I.~Joumard, S.~Blanchard, J.~Marcus, R.~Cubitt, T.~Giamarchi, and 
   P.~Le~Doussal, 
   Nature {\bf 413}, 404 (2001).

\bibitem{Kierfeld}
   J.~Kierfeld, T.~Nattermann, and T.~Hwa, 
   Phys. Rev. B {\bf 55}, 626 (1997).

\bibitem{FisherD}
   D.S.~Fisher, 
   Phys. Rev. Lett. {\bf 78}, 1964 (1997). 

\bibitem{Menon}
   G.I.~Menon, 
   Phys. Rev. B {\bf 65}, 104527 (2002).  

\bibitem{Du}
   X.~Du, G.~Li, E.Y.~Andrei, M.~Greenblatt, and P.~Shuk,
   Nature Physics {\bf 3}, 111 (2007).

\bibitem{Henderson}
   W.~Henderson, E.Y.~Andrei, M.J.~Higgins, and S.~Bhattacharya,
   Phys. Rev. Lett. {\bf 77}, 2077 (1996).

\bibitem{Nicodemi}
   M.~Nicodemi and H.J.~Jensen, 
   Phys. Rev. Lett. {\bf 86}, 4378 (2001);
   J. Phys. A: Math. Gen. {\bf 34}, L11 (2001);
   Europhys. Lett. {\bf 54}, 566 (2001); 
   J. Phys. A: Math. Gen. {\bf 34}, 8425 (2001); 
   Phys. Rev. B {\bf 65}, 144517 (2002).

\bibitem{Olson}
   C.J.~Olson, C.~Reichhardt, R.T.~Scalettar, G.T.~Zimanyi, and 
   N.~Gr{\o}nbach-Jensen, 
   Phys. Rev. B {\bf 67}, 184523 (2003).

\bibitem{BCD1}
   S.~Bustingorry, L.F.~Cugliandolo, and D.~Dom\'inguez, 
   Phys. Rev. Lett. {\bf 96}, 027001 (2006).

\bibitem{BCD2}
   S.~Bustingorry, L.F.~Cugliandolo, and D.~Dom\'inguez, 
   Phys. Rev. B {\bf 75}, 024506 (2007).

\bibitem{BCI}
   S.~Bustingorry, L.F.~Cugliandolo, and J.L.~Iguain, 
   J. Stat. Mech. {\bf (2007)} P09008.

\bibitem{IBKC} 
   J.L.~Iguain, S.~Bustingorry, A.B.~Kolton, and L.F.~Cugliandolo, 
   Phys. Rev. B {\bf 80}, 094201 (2009).

\bibitem{Roma}
   F. Rom\'a and D. Dom\'inguez, 
   Phys. Rev. B {\bf 78}, 184431 (2008).

\bibitem{Otterlo}
   A. van Otterlo, R.T.~Scalettar, and G.T.~Zim\'{a}nyi,  
   Phys. Rev. Lett. {\bf 81}, 1497 (1998).
   
\bibitem{Vinokur}
   D.R.~Nelson and V.M.~Vinokur, 
   Phys. Rev. B {\bf 48}, 13 060 (1993).

\bibitem{Sen}
   P.~Sen, N.~Trivedi, and D.M.~Ceperley, 
   Phys. Rev. Lett. {\bf 86}, 4092 (2001).

\bibitem{Rosso}
   A.~Rosso and W.~Krauth, 
   Phys. Rev. B {\bf 65}, 012202 (2001)

\bibitem{Tyagi}
   S.~ Tyagi and Y.Y.~Goldschmidt, 
   Phys. Rev. B {\bf 67}, 214501 (2003).

\bibitem{Petaja}
   V.~Pet\"aj\"a, M.~Alava, and H.~Rieger, 
   Europhys. Lett. {\bf 66}, 778 (2004).

\bibitem{Bullard}
   J.~Das, T.J.~Bullard, and U.C.~T\"auber, 
   Physica A {\bf 318}, 48 (2003); 
   T.J.~Bullard, J.~Das, G.L.~Daquila, and U.C.~T\"auber,
   Eur. Phys. J. B {\bf 65}, 469 (2008).

\bibitem{Thananart}
   T.~Klongcheongsan, Ph.D. thesis, Virginia Tech (2009);
   T.~Klongcheongsan, T.J.~Bullard, and U.C.~T\"auber, 
   Supercond. Sci. Technol. {\bf 23}, 025023 (2010).

\bibitem{Rothlein}
   A.~R\"{o}thlein, F.~Baumann, and M.~Pleimling, 
   Phys. Rev. E {\bf 74}, 061604 (2006); 
   Phys. Rev. E {\bf 76}, 019901(E) (2007).

\bibitem{Bustingorry07} 
   S.~Bustingorry, 
   J. Stat. Mech. {\bf (2007)} P10002.

\bibitem{Chou10} 
   Y-L.~Chou and M.~Pleimling, 
   J. Stat. Mech. {\bf (2010)} P08007.

\bibitem{Edw82} 
   S.F.~Edwards and D.R.~Wilkinson, 
   Proc. R. Soc. London Ser. A {\bf 381}, 17 (1982).

\bibitem{Noh09} 
   J.D.~Noh and H.~Park, 
   Phys. Rev. E {\bf 80}, 040102(R) (2009).

\bibitem{Mon09} 
   C.~Monthus and T.~Garel, 
   J. Stat. Mech. {\bf (2009)} P12017.

\bibitem{Par10}
   H.~Park and M.~Pleimling, 
   Phys. Rev. B {\bf 82}, 144406 (2010).

\bibitem{Cor10}
   F.~Corberi, L.F.~Cugliandolo, and H.~Yoshino, 
   Preprint {\tt arXiv:1010.0149}.

\bibitem{Cor11}
   F.~Corberi, E.~Lippiello, A.~Mukherjee, S.~Puri, and M.~Zannetti, 
   J. Stat. Mech. {\bf (2011)} P03016.

\bibitem{Gotze}
  W.~G\"otze and L.~Sjogren,
  Rep. Prog. Phys. {\bf 55}, 241 (1992).

\bibitem{Dob11}
   U.~Dobramysl, M.~Pleimling, and U.C.~T\"auber 
   (work in progress).

\end{thebibliography}
\end{document}